\documentclass[12pt]{article}
\usepackage[cm]{fullpage}
\usepackage{authblk} %affil.
\usepackage{graphicx}
\graphicspath{ {./} }
\usepackage{hyperref}

\begin{document}

\title{A Brief Review of Plasma Wakefield Acceleration}

\author{Altan Cakir\footnote{Corresponding author\\ \textit{Email addresses:} \textbf{\href{mailto: altan.cakir@itu.edu.tr}{altan.cakir@itu.edu.tr}} (Altan Cakir), \textbf{\href{guzelah@itu.edu.tr}{guzelah@itu.edu.tr}} (Oguz Guzel) \\ \textit{Submitted to ..............}} \space and Oguz Guzel}

\affil{\small{\textit{Department of Physics Eng., Istanbul Technical University, 34469, Istanbul, Turkey}}}

\maketitle

\hrulefill

\abstract{Plasma Wakefield Accelerators could provide huge acceleration gradients that are 10 - 1000 times greater than conventional radio frequency metallic cavities available in current accelerators and at the same time the size of plasma wakefield accelerators could be much smaller than today's most succesful colliders. This review gives brief explanations of the working principle of Plasma Wakefield Accelerators and shows the recent development of the field. The current challenges are given and the potential future use of Plasma Wakefield Accelerators are discussed.}

\begin{flushleft}
	\textbf{Keywords:} plasma wakefield accelerator; laser wakefield acceleration, beam-driven plasma wakefield acceleration\\
\end{flushleft}

\hrulefill

\section{Introduction}

Since the rise of the particle physics in the 20th century, the particle accelerators became crucial to further grasp the fundamental theories of the particle physics. Today, the particle accelerators are state-of-the-art technology and are able to test the governing forces and the interactions between very tiny fractures of the visible matter. The current milestone, with a circumference of 27 kilometers is the Large Hadron Collider (LHC) at the European Organization for Nuclear Research (CERN), situated in the Franco-Swiss border. The LHC accelerates proton beams to nearly the speed of light gaining them an energy of 6.5 TeV. Today, the LHC is the biggest particle accelerator in the world and is financed by numerous countries across the Europe. A bigger accelerator, Future Circular Collider (FCC), is being designed. A conceptual design report for the FCC was submitted in December 2018 stating that the FCC is planned to be hosted in a 100 kilometers-long tunnel \cite{Abada2019}.

The reason for constructing bigger accelerators is that the acceleration gradient is proportional to the distance. Therefore, the main limit for the acceleration of particles in linacs (linear accelerator) is the dimension of the accelerator itself and the field strength of the bending magnets and energy loss through snycrotron radiate -which scales as $ E^{4} $ - in circular accelerators. It is clear that the conventional accelerators are becoming very big and costly \cite{CERN-YR214}. Novel accelerator schemes must be investigated to overcome this strict size-energy relation.

Acceleration of the charged particles inside a plasma medium is a viable solution to the problem. At small distances, the plasmas can afford high electric field gradients of about $ 10-100 GV m^{-1} $ whereas the conventional RF accelerators require several kilometers of long tunnels to acquire the same gradient \cite{PhysRevLett.43.267}. The main drawback here is that the plasma wakefield accelerators (PWFA) cannot have an infinetely long cell length, due to phase slippage and difficulties maintaining a uniform plasma density to be considered as a replacement for today's accelerators. Nevertheless, this feature gives us the opportunity to shrink the dimension of the future accelerators in a large scale. The encouraging experimental results that are achieved so far show in the near future, the energy frontier can be extended and the accelerators may become more compact structures or table-top experiments. Today, the hopeful potential of plasma-based acceleration scheme is regarded as one  of the best future replacements of the current RF accelerators.

\section{Laser Wakefield Acceleration}

\begin{figure}[h!]
	\centering
	\includegraphics[width=10cm]{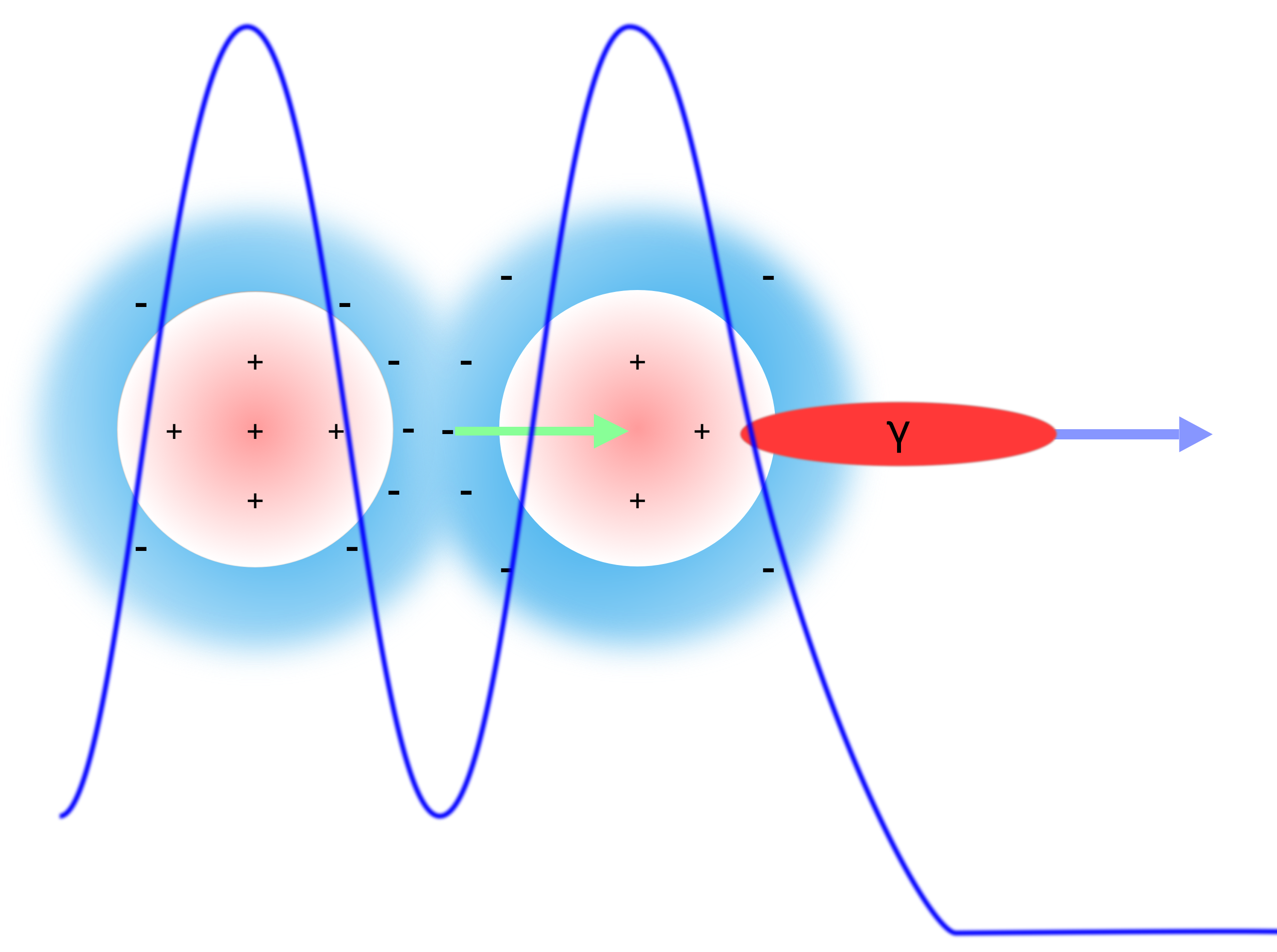}
	\caption{Schematic of the LWFA. A laser pulse (red) when injected into a plasma medium, knocks away the plasma electrons (blue) and creates wakefields (pink) which are co-moving with the laser beam at a speed close to the speed of light. The electrons behind are pulled towards by the positively charged wakefield hence accelerated. The blue line represents the ionization level of the plasma.}
	\label{fig:lwfa-drawing}
\end{figure}

The first idea of creating high electric field gradients inside a plasma medium by a laser pulse was proposed by Tajima and Dawson in 1979 \cite{PhysRevLett.43.267}. Their proposed mechanism involved sending a short laser pulse through a plasma. The radiation pressure of an intense laser pulse displaces the plasma electrons ahead and aside creating a positively charged ion wake, namely a wakefield behind and co-moving with the laser pulse. After the passage of the laser pulse, the displaced electrons are pulled back and caught by the positively charged wakefield and are accelerated longitudinally (see Figure \ref{fig:lwfa-drawing}). This type of particle acceleration came to be known as the laser wakefield acceleration (LWFA).

The most efficent way of driving a wakefield and producing quasi-monoenergetic electron beams by a laser can be performed when the laser pulse length is on the order of the plasma wavelength \cite{doi:10.1109/27.509991}, $ c / w_{p} $ where,

\begin{equation}
\label{eq:1}
	w_{p} =  \sqrt{\frac{e^{2} n_{e}}{m_{e} \epsilon_{0}}} ,
\end{equation}

\noindent $ n_{e} $ is the number density of the plasma electrons in $ cm^{-3} $, $\epsilon_{0}$ is the permittivity of free space, e and $  m_{e} $ are the electron's charge and rest mass . This regime is called the "bubble" regime and the most of the PWFA experiments today are in this regime \cite{CERN-YR214}. In the bubble regime the energy of the laser pulse is concentrated inside a sphere of radius smaller than the radius of the plasma wakefield. In the sphere, the ponderomotive force\footnote{The electromagnetic force exerted on a charged particle in an oscillating electromagnetic field.} radially knocks away the plasma electrons then the wakefields are created.

The electric field E, that a plasma can support is given by the following relation \cite{RevModPhys.81.1229}:

\begin{equation}
\label{eq:2}
E = \frac{c m_{e} w_{p}}{e} .
\end{equation} 
 
\noindent Substituting $ w_{p} $ in the Eq. \ref{eq:1} to Eq. \ref{eq:2} shows that the electric field is proportional to $ n_{e}^{1/2} $. Tajima and Dawson estimated that a laser of wavelength 1 $ \mu m $ shone on a plasma of density $ 10^{18}$  $ cm^{-3} $ could accelerate electrons with an electric field gradient of 10 $ GeVm^{-1} $ \cite{PhysRevLett.43.267}. This gradient is 3 orders of magnitude greater than the acceleration gradients of the today's linear RF accelerators.

\begin{figure}[h!]
	\centering
	\includegraphics[width=15cm]{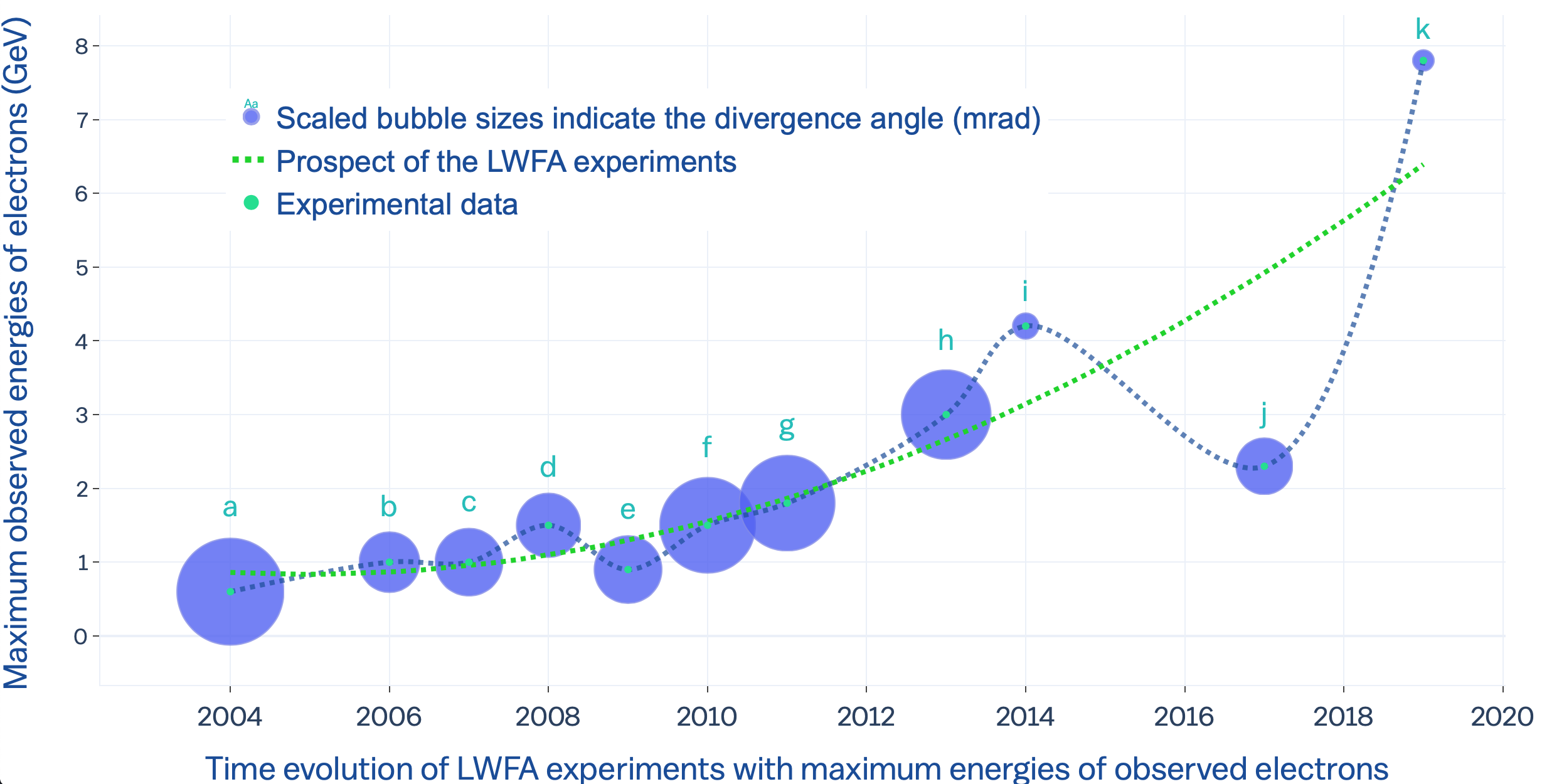}
	\caption{Achieved energies of electrons in LWFA experiments according to years.}
	\label{fig:lwfa-figure}
\end{figure}

Throughout the development of the lasers and the relevant technological equipment, the LWFA became one of the most promising alternative acceleration techniques. Fig. \ref{fig:lwfa-figure}. shows the time evolution of the observed energy of the electrons in LWFA experiments around the world. We added the important experimental data after 2004 since the LWFA experiments before 2004 yielded 100\% energy spreads \cite{Mangles2004}. Figure \ref{fig:lwfa-figure} clearly indicates the success of the PWFA experiments on the divergence angle decrease throughout the years. This simply means that the laser wakefield accelerators can produce mono-energetic electron beams with very less energy spread. However, in the case of Kim et al. in 2013 \cite{Kim2013} and in 2017 \cite{Kim2017}, the decrease in the divergence angle yielded the decrease in the electric field gradient and also in the maximum energy. Nevertheless, the LWFA today is capable of producing electrons of energies on the order of 10 GeV. It is clear that the technological development of the lasers pushed the acceleration of the electrons to the higher energies.

\begin{table}[h!]
	\caption{Tabular data of Figure \ref{fig:lwfa-figure}.}
	\centering
	\begin{small}
	
		\begin{tabular}{clcccc}
				
		\hline
		\hline
		& Experiment               & Year & \multicolumn{1}{c}{Achieved Energy} & \multicolumn{1}{c}{Acceleration Gradient} & \multicolumn{1}{c}{Divergence Angle} \\
		
		&&&(GeV)&(GeV/m)&(mrad)\\
		
			\hline
			\\

			a &	Faure, J. et al. \cite{Faure2004}          & 2004 & 0.17                  & 57                         & 10                      \\
			b &Leemans, W. P. et al. \cite{Leemans2006}    & 2006 & 1                     & 30                          & 1.6                    \\
			c &Karsch, et al. \cite{Karsch2007}             & 2007 & 1                     & 66                   & 2                       \\
			d &Hafz, N. A. M. et al. \cite{Hafz2008}    & 2008 & 1.5                   & 300                           & 1.8                    \\
			e &Froula, D. H. et al. \cite{Froula2009}     & 2009 & 0.72                  & 90                            & 2.85                   \\
			f &Clayton, C. E. et al. \cite{Clayton2010}    & 2010 & 1.45                  & 111                        & 4.4                    \\
			g &Lu, H. et al.  \cite{Lu2011}           & 2011 & 1.8                   & 45                            & 4.5                    \\
			h &Kim, H. T. et al. \cite{Kim2013}  & 2013 & 3                     & 300                           & 4                       \\
			i &BELLA, (2014) \cite{Leemans2014}             & 2014 & 4.2                   & 46                         & 0.3                     \\
			j &Kim, H. T. et al. \cite{Kim2017}  & 2017 & 2.3                   & 230                           & 1.4                    \\
			k &BELLA, (2019) \cite{PhysRevLett.122.084801}             & 2019 & 7.8                   & 86                         & 0.2

\end{tabular}
\end{small}
\end{table}

The laser technology was not able to produce such short laser pulses at that time, therefore as an alternative, the Plasma Beatwave Acceleration (PBWA) was proposed by Tajima and Dawson. They noted that the plasma wakefields can also be excited by two long laser pulses with slightly different frequencies. The wavelengths of the two lasers are arranged to be equal to the plasma wavelength and to resonantly excite a plasma wakefield. Afterwards, the plasma electrons are accelelerated by the same scheme as in the LWFA. This method of the plasma wakefield acceleration came to be known as the plasma beatwave acceleration (PBWA) \cite{doi:10.1109/27.509991}. The first observation of the PBWA was done by a group at the University of California, Los Angeles (UCLA) in 1985 \cite{PhysRevLett.54.2343} and later on by many groups \cite{PhysRevLett.70.37,ebrahim,dangor,PhysRevLett.68.48,PhysRevLett.68.3710,Ebrahim1994,doi:10.1063/1.870730,AIPConferenceProceedings.335.612}.

However, the PBWA had a challenging limit. The saturation of the plasma wakefield amplitude occurs since the increase in mass of relativistic electrons decreases the plasma frequency breaking the beatwave resonance (see equation \ref{eq:1}) \cite{PhysRevLett.29.701}. The consequence of this effect was the tendency of the researchers to the LWFA.

A challenging side was the energy spread at the time when the LWFA is proposed. 
Early experiments of the LWFA were producing electron beams with broad energy spread, however, with the application of the chirped pulse amplication to lasers by Strickland and Mourou \cite{STRICKLAND1985447,Mourou1998} which is awarded the Nobel Prize in Physics in 2018, the LWFA was shown to be able to produce mono-energetic electron beams using short laser pulses. The first experiments using this technology were conducted by three independent groups in 2004 \cite{Faure2004,doi:10.1038/nature02900,Mangles2004}. This  played a crucial role in the development of the laser wakefield accelerators; the wakefields became more bubble-shaped with one-two periods rather than being a blur density change with multiple periods \cite{CERN-YR214}.

Today, many groups are doing experiment on LWFA achieving the GeV energies \cite{PhysRevLett.103.035002,Leemans2014,Hafz2008}. The current milestone, in January 2019, is achieved by accelerating electrons to 7.8 GeV energy at the Berkeley Lab Laser Accelerator (BELLA). BELLA experiment nearly doubled the energy record for the LWFA achieved in 2014 at the same laboratory \cite{PhysRevLett.122.084801}. 

Another technique of accelerating the plasma electrons by using the lasers is called the self-modulated LWFA (SM-LWFA). This type of the LWFA is studied theoretically for the first time by Krall et al. and Andreev et al. in the first half of the 1990s \cite{PhysRevE.48.2157,PhysRevLett.74.4428}. The difference of the scheme is that the laser pulse length and the plasma density are greater than the ones in LWFA. The laser pulse is modulated at the plasma frequency because of the interaction of the pulse with the plasma. The mechanism for the modulation is the self-focusing and the defocusing. Self-focusing is a phenomenon such that a non-linear medium whose index of refraction is inversely proportional to the local electric field intensity, focuses the electromagnetic waves that travel through and the defocusing works vice versa \cite{PhysRevLett.60.1298}. The modulated laser pulse triggers the formation of wakefields as same as in the PBWA scheme. The first evidence of plasma wakefield formation in the self-modulated regime was observed by Coverdale et al. in 1995 \cite{PhysRevLett.74.4659}. Afterwards, the SM-LWFA was shown to be capable of accelerating the electrons \cite{PhysRevLett.74.4428,Modena1995}.

\section{Beam Driven Plasma Wakefield Acceleration}

The idea of accelerating the charged particles with the plasma waves driven by the beams of particles was first given by Fainberg et al. in 1956 \cite{fainberg1056} and the theory of PWFA in the linear regime was proposed by Chen et al. in 1985 \cite{PhysRevLett.54.693}. The proposed mechanism was to inject a sequence of high-energetic electron beams (driver beam) into a plasma medium. The driver beam knocks away the plasma electrons and creates a positively charged wakefield that co-moves with and behind the leading electron beam. A second injected electron beam (witness beam) with a proper phase is accelerated longitudinally to high energies by the electrical attraction of the positively charged wakefield. The early experiments with relativistic electrons verified the proposal \cite{PhysRevLett.61.98,doi:10.1063/1.859559,doi:10.1063/1.44061}. This type of plasma-based acceleration came to be known as the Beam-Driven Plasma Wakefield Acceleration (BD-PWFA).

The first experimental observation of BD-PWFA was shown at the Argonne National Laboratory's Advanced Accelerator Test Facility in 1988 \cite{PhysRevLett.61.98}. They achieved the $ MVm^{-1} $ order of electric field gradient by injecting 21 MeV-energetic driver electron bunch into a plasma of density $ n_{e} = 10^{13} cm^{-3}$. In the experimental setup, a low intensity witness pulse (electrons) of 15 MeV is injected at a proper delay time with the high intensity driver bunch. There are many other experiments of electron-driven PWFA (see Table \ref{table:PWFA}). The current milestone is the SLAC's experiment in 2007 \cite{Blumenfeld2007}. They used 42 GeV-energetic electron beams to drive wakefields inside a plasma medium $ 2.7 \times 10^{17} cm^{-3} $ of density. The outcome was the doubled energy of the incoming electrons and an acceleration gradient of about 52 $ GeVm^{-1} $.

In 2003, Blue et al. showed for the first time the formation of wakefields driven by a positron beam at the Stanford Linear Accelerator Center (SLAC) \cite{PhysRevLett.90.214801}. Contrary to the electron driven PWFA, a positron beam when injected into a plasma medium, attracts the plasma electrons by the transverse electric field. The positrons first decelerated due to the work done on the plasma electrons however, once the attracted electrons pass the propagation axis of the positron beam, the transverse electric field changes sign to positive and pushes the positrons. Blue et al. used 28.5 GeV-energetic and 2.4 ps long positron beam into a $ 1.8 \times 10^{14} cm^{-3} $ of electron density of the plasma. In the experiment, an acceleration gradient of $ \sim 56 MeV/m$ is achieved inside the 1.4 meters of lithium plasma. Another experiment on positron-driven PWFA was conducted by Corde et al. in 2015 \cite{Corde2015}. They showed a 5 GeV energy gain of the positrons over a distance of 1.3 meters which corresponds to an acceleration gradient of about 3.8 $ GeVm^{-1} $. There is also another experimental setup which is using a laser beam as a driver to accelerate positrons \cite{Gessner2016}. These hopeful experimental results are regarded to be important for the plasma-based linear colliders that are thought to be designed and built in the close future.

\begin{figure}[h!]
	\centering
	\includegraphics[width=15cm]{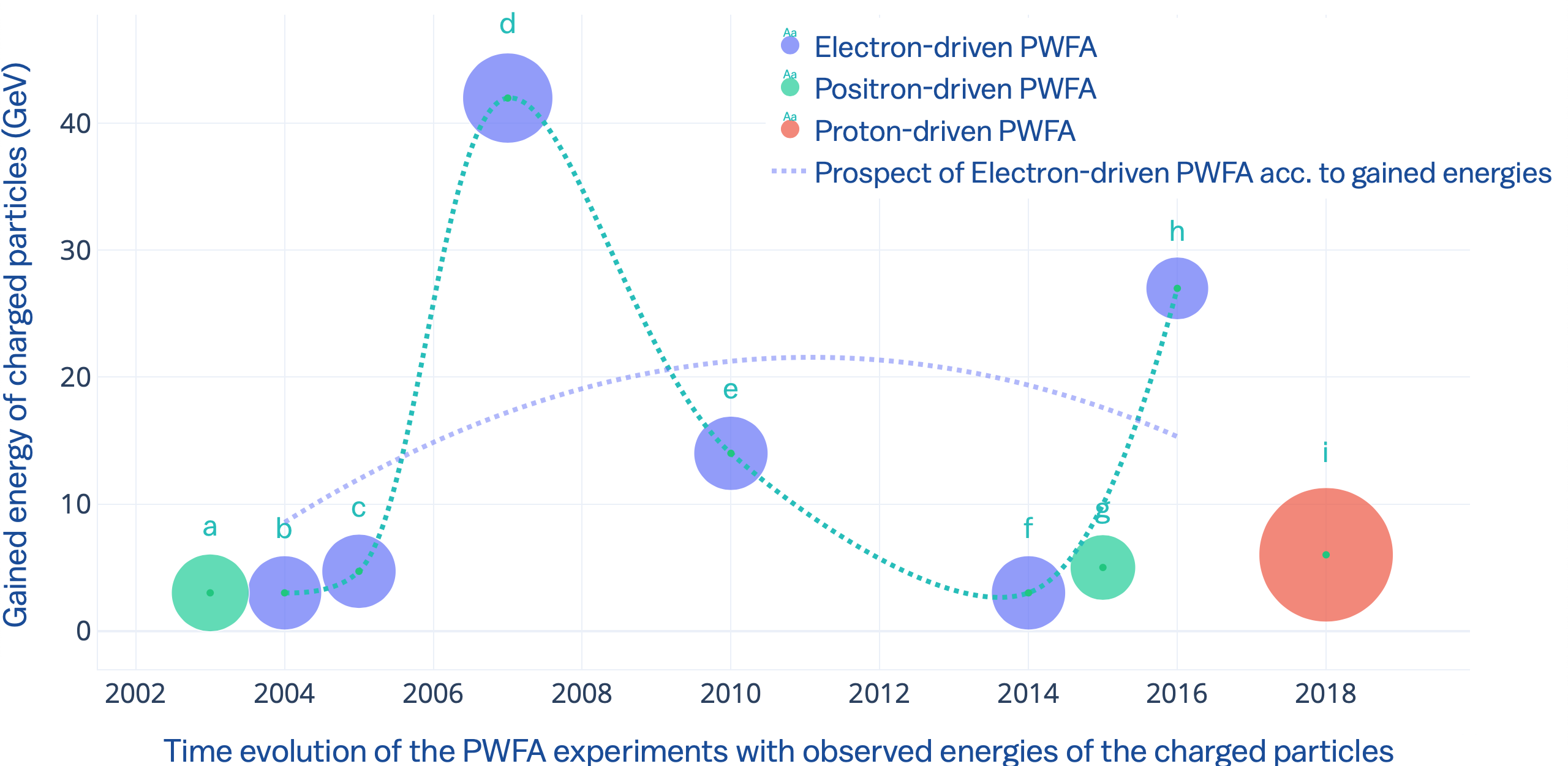}
	\caption{Achieved energies of electrons in PWFA experiments according to years.}
	\label{fig:pwfa-figure}
\end{figure}

\begin{table}[h!]
	\caption{Tabulated data of Figure \ref{fig:pwfa-figure}}
	\label{table:PWFA}

	\begin{small}

	\begin{tabular}{llccccc}
		\hline
		\hline
		& Experiment            & Year & Energy Gain & Acceleration Gradient & Driver Type & Driver Energy\\
		& & &(GeV)&(GeV/m)& &(GeV)\\
		\hline
		\\
		a & Blue, B. E. et al. \cite{Blue2003}   & 2003 & 0.08              & 0.056                         & positron    & 28.5                \\
		b & Muggli, P. et al.  \cite{Muggli2004}   & 2004 & 0.28              & 0.2                           & electron    & 28.5                \\
		c & Hogan, M. J. et al. \cite{Hogan2005}  & 2005 & 2.7               & 27                            & electron    & 28.5                \\
		d & Blumenfeld, I. et al. \cite{Blumenfeld2007}    & 2007 & 42                & 52                            & electron    & 42                  \\
		e & Muggli, P. et al. \cite{Muggli2010}        & 2010 & 14                & 36                            & electron    & 28.5                \\
		f & Litos, M. et al. \cite{Litos2014} & 2014 & 1.6               & 4.4                           & electron    & 28.5                \\
		g & Corde, S. et al. \cite{Corde2015} & 2015 & 5				& 1.3 							& positron		&	20.35 \\
		h & Corde, S. et al. \cite{Corde2016}     & 2016 & 27                & 150                           & electron    & 20.35               \\
		i & Adli, E. et al. \cite{Adli2018}      & 2018 & 2                 & 0.2                           & proton      & 400         
		       
	\end{tabular}
	\end{small}
\end{table}

The proton-driven PWFA is another scheme of this developing accelerating technique and is very promising since the high-energetic protons are already available in the RF accelerators and have a potential to provide greater acceleration gradients. The first experimental setup for proton-driven PWFA, namely the Advanced Wakefield Acceleration Experiment (AWAKE) was developed at the CERN and its first run took place between 2016 and 2018 \cite{Gschwendter:2019uic}. The experiment uses the long proton bunch of 400 GeV energy supplied from the CERN's Super Proton Synchrotron (SPS). A laser pulse is injected together with the proton bunch to create an ionization front in the 10 meters of rubidium vapor and to seed the self-modulation of the long proton bunch into microbunches. The self-modulation of a long particle bunch is such a phenomenon that it can be triggered by the instabilities in the plasma \cite{Caldwell:2017xme}. In BD-PWFA the acceleration gradient is proportional to the $ N/\sigma^{2} $, where N is the number of particles of the driving beam and $\sigma$ is the bunch length \cite{Joshi2002}. The advantage of self-modulation here is that, decrasing the bunch length yield higher electric field gradients and is much more efficient than increasing the number of the particles in the bunch. 10-20 MeV-energetic electrons are injected in between these proton microbunches to investigate the wakefields and the acceleration. Finally, the trapped electrons are pushed to higher energies by the following proton microbunch. The AWAKE experiment succesfully accelerated the externally injected electrons to 2 GeV in this scheme \cite{Adli2018}. 

Figure \ref{fig:pwfa-figure} shows the major experiments around the world demonstrating the BD-PWFA and the Table \ref{table:PWFA} shows the corresponding data. The BD-PWFA experiments using protons and positrons are very rare and high acceleration gradients on the order of 10 GeV have not been seen yet. However, the electron driven PWFA shows greate promise although there has been a decrease between 2007 and 2014.

There are many other experiments/laboratories studying PWFA such as MAX IV Laboratory (Sweden) \cite{maxiv}, Central Laser Facility - Rutherford Appleton Laboratory (UK) \cite{clf}, FACET II Experiment - SLAC (USA) \cite{facet-ii}, LUX Group - Hamburg University (Germany) \cite{lux}, CILEX - APOLLON Facility (France) \cite{cilex}.

\section{The challenges and the future outlook}

The plasma wakefield accelerator scheme seems as one of the best options that can be the future collider of the particle physics. However there are several challenges that must be overcome in advance. For the LWFA scheme, the laser systems have the capability of providing PW power but they are still weak in pulse repetition rate. In addition, these PW laser systems have not yet achieved wall-plug efficiency, which is a big obstacle for the plasma wakefield accelerators, to become the table-top experiments. For an electron-positron plasma wakefield collider, the acceleration scheme for the positrons must be developed since the PWFA works better for the electrons. This is due to the decelerating force of the plasma electrons exerted on the positrons. Moreover, in order to accept the accelerated particles as suitable for colliding, they must be stable while being accelerated in the plasma medium, on top of that, their energy spread must be low. To overcome some of these problems, an interesting acceleration scheme is proposed, namely Hybrid Laser-Plasma Wakefield Accelerator \cite{PhysRevLett.104.195002}. Also an $ e^{-} $ $ e^{+} $ plasma wakefield collider is being designed addressing some of these issues \cite{MUGGLI2009116}.

The TeV plasma wakefield accelerator has much more challenging requirements. The current energy level for the plasma wakefield accelerators is about tens of GeV. The next one will be the 100 GeV electrons and requires greater laser pulse energy or several laser pulses. Besides, multi stages of acceleration or a single long stage need to be applied to reach TeV energies and this implies the longer total length of the plasma wakefield accelerator. The TeV energies seem not so close in the future by PWFA. The plasma wakefield accelerators seem also the most suitable future compact sources of radiation. Its potential application areas include radiography of materials, radiotherapy and radiolysis in biology \cite{malka,gauduel}.\\

\bibliographystyle{unsrt}
\bibliography{references.bib}

\end{document}